\documentclass[a4paper]{article}

\usepackage{ISCSLP2024}
\usepackage{graphicx}
\usepackage{subfig}

\title{EELE: Exploring Efficient and Extensible LoRA Integration in Emotional Text-to-Speech}
\name{Xin Qi$^{1,2}$,  Ruibo Fu$^{*1}$, Zhengqi Wen$^5$, Jianhua Tao$^{4,5}$, Shuchen Shi$^6$, Yi Lu$^{1,2}$, Zhiyong Wang$^{1,2}$, Xiaopeng Wang$^{1,2}$, Yuankun Xie$^3$,  Yukun Liu$^1$, Guanjun Li$^1$, Xuefei Liu$^1$, Yongwei Li$^1$}
\address{
 $^1$Institute of Automation, Chinese Academy of Science\\
  $^2$School of Artificial Intelligence, University of Chinese Academy of Sciences\\
  $^3$School of Information and Communication Engineering, Communication University of China\\
  $^4$Department of Automation, Tsinghua University\\
  $^5$Beijing National Research Center for Information Science and Technology, Tsinghua University\\
  $^6$Shanghai Polytechnic University, China
  }
\email{qixin221@mails.ucas.ac.cn, ruibo.fu@nlpr.ia.ac.cn}

\begin{document}

\maketitle

\renewcommand{\thefootnote}{} 
\footnotetext[1]{* denotes corresponding author.}
\begin{abstract}
In the current era of Artificial Intelligence Generated Content (AIGC), a Low-Rank Adaptation (LoRA) method has emerged. It uses a plugin-based approach to learn new knowledge with lower parameter quantities and computational costs, and it can be plugged in and out based on the specific sub-tasks, offering high flexibility. However, the current application schemes primarily incorporate LoRA into the pre-introduced conditional parts of the speech models. This fixes the position of LoRA, limiting the flexibility and scalability of its application. 
Therefore, we propose the Exploring Efficient and Extensible LoRA Integration in Emotional Text-to-Speech (EELE) method. Starting from a general neutral speech model, we do not pre-introduce emotional information but instead use the LoRA plugin to design a flexible adaptive scheme that endows the model with emotional generation capabilities. Specifically, we initially train the model using only neutral speech data. After training is complete, we insert LoRA into different modules and fine-tune the model with emotional speech data to find the optimal insertion scheme. Through experiments, we compare and test the effects of inserting LoRA at different positions within the model and assess LoRA's ability to learn various emotions, effectively proving the validity of our method. Additionally, we explore the impact of the rank size of LoRA and the difference compared to directly fine-tuning the entire model.
\end{abstract}

\begin{figure*}[!t]
	\centering
	\subfloat[Text information modeling]{\includegraphics[width=0.25\linewidth]{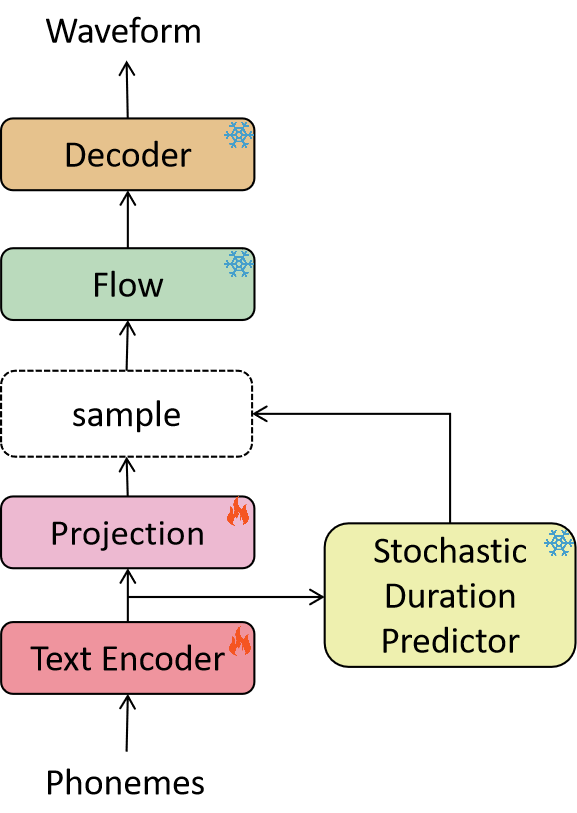}
		\label{fig_first_case}}
	\subfloat[Distribution transformation]{\includegraphics[width=0.25\linewidth]{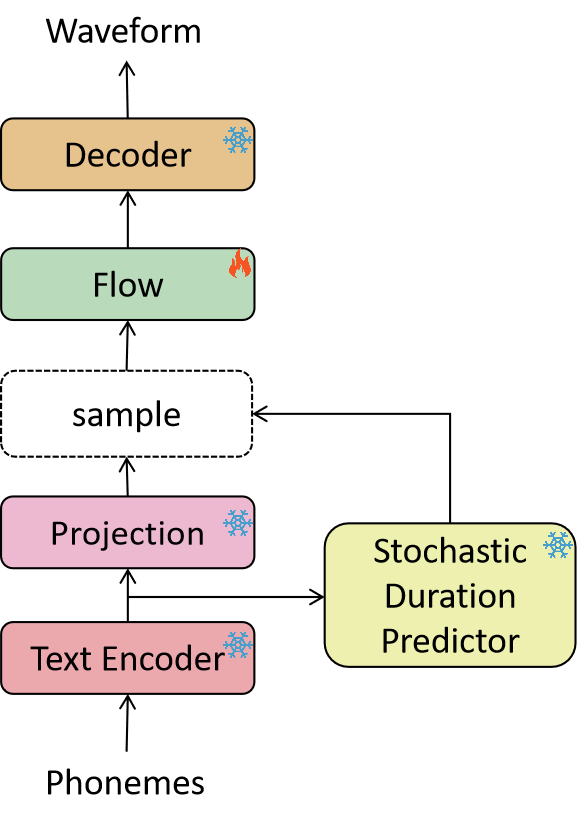}
		\label{fig_first_case}}
	\subfloat[Decode]{\includegraphics[width=0.25\linewidth]{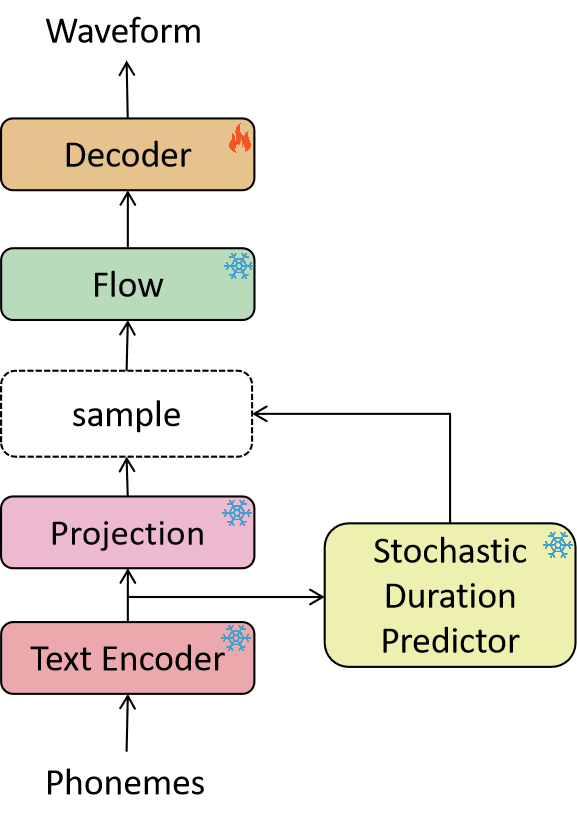}
		\label{fig_first_case}}
	\subfloat[Acoustic information modeling]{\includegraphics[width=0.25\linewidth]{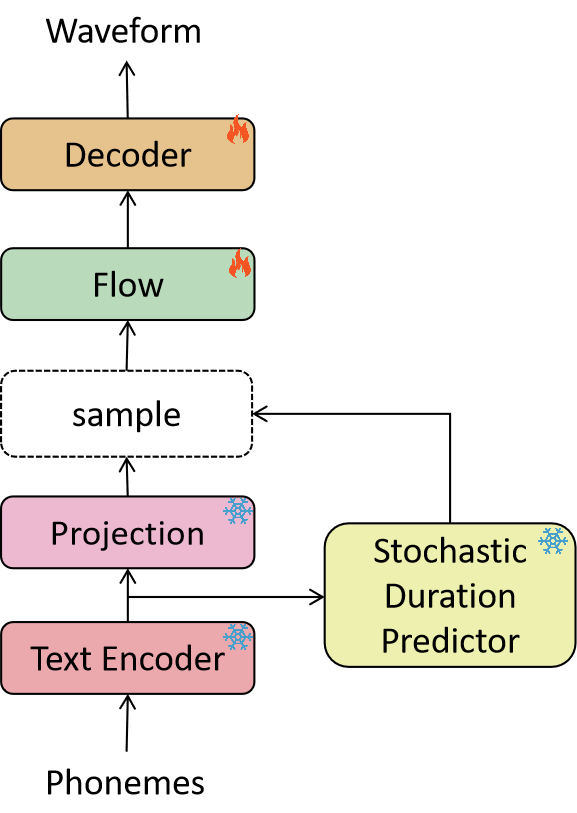}
		\label{fig_first_case}}

	\subfloat[Duration ]{\includegraphics[width=0.25\linewidth]{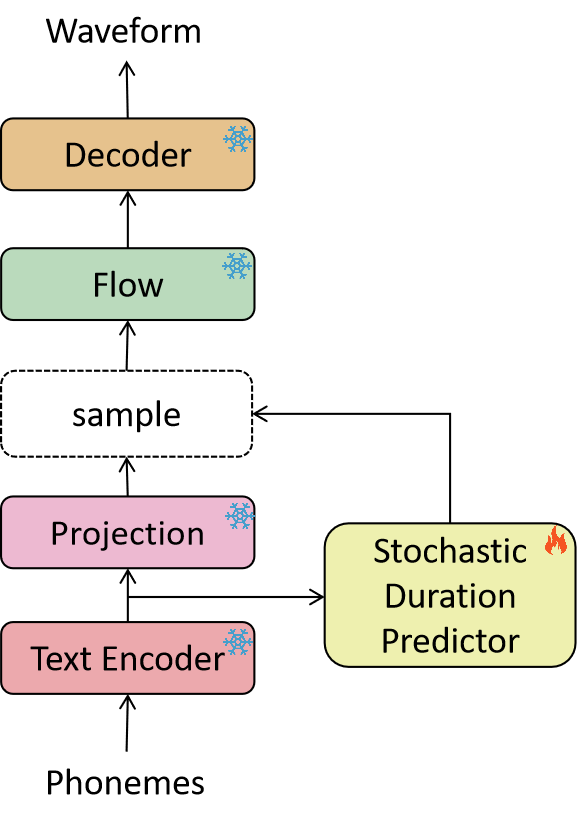}
		\label{fig_first_case}}
	\subfloat[Duration and text information]{\includegraphics[width=0.25\linewidth]{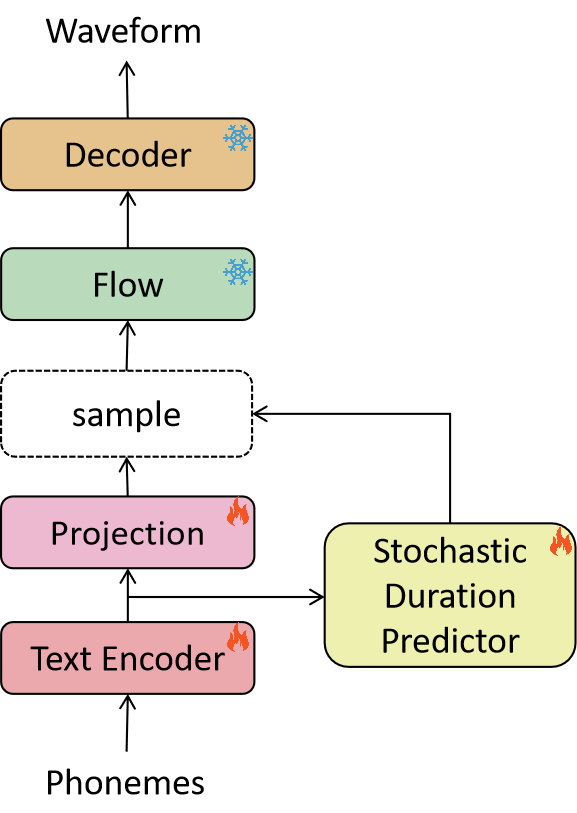}
		\label{fig_first_case}}
    \subfloat[Duration and acoustic information]{\includegraphics[width=0.25\linewidth]{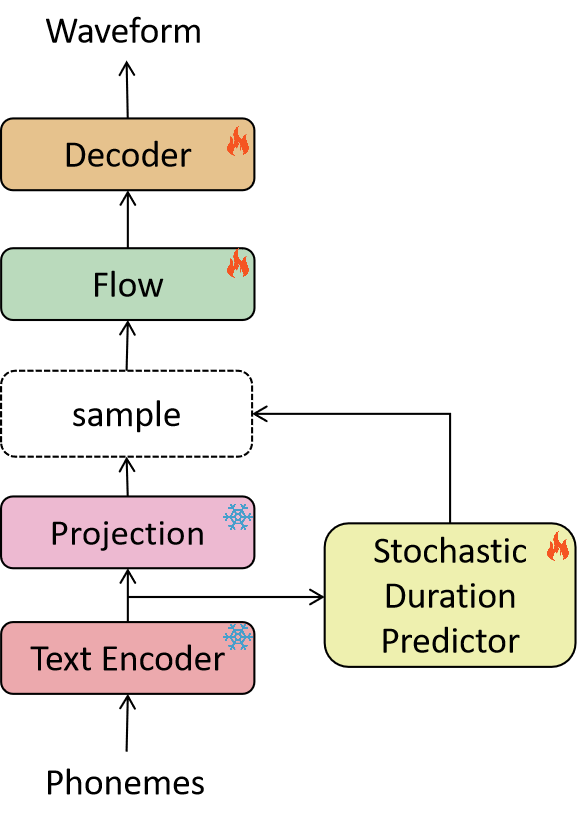}
		\label{fig_first_case}}
	\subfloat[Duration, acoustic information and projection]{\includegraphics[width=0.25\linewidth]{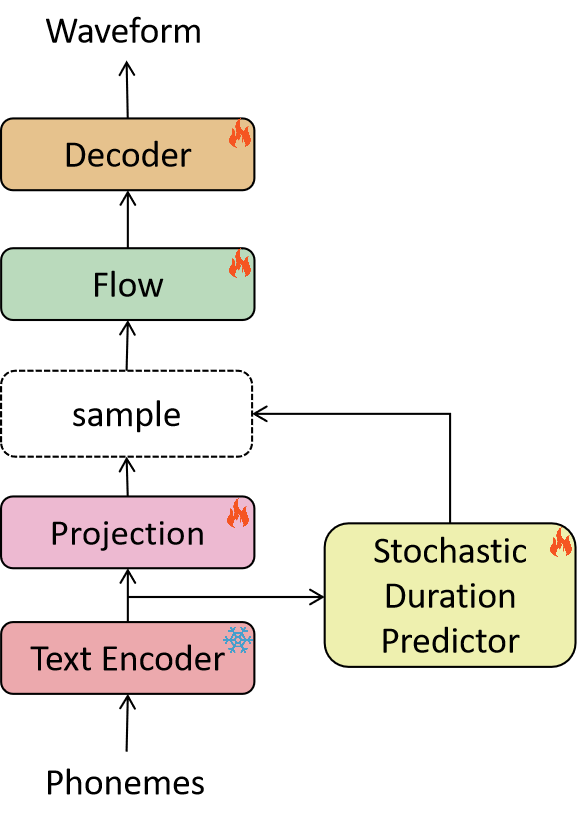}
		\label{fig_first_case}}
	\caption{An overview of all attempts to add LoRA to the model. \includegraphics[width=1em]{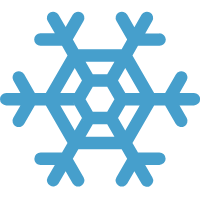} represents freezing all parameters. \includegraphics[width=1em]{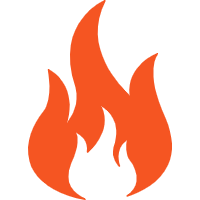} represents freezing all parameters and adding LoRA for fine-tuning.}
	\label{all_try}
\end{figure*}
\section{Introduction}
The development of text-to-speech (TTS) has reached a point where it can generate very high-quality speech. 
For example, VITS \cite{kim2021conditional} and Grad-TTS \cite{popov2021grad} have incorporated complex neural network architectures and training strategies to generate results that are very close to real speech.
In this field, emotion generation has always been a hot topic that researchers are constantly exploring.
Adding emotion to speech not only enhances the realism of generated results, but also has significant application value in various real-life scenarios, such as in movies, games, and voice acting.

There has been a significant amount of work on emotion TTS, but most of these approaches incorporate emotional information during the initial model training stage.
These approaches can be broadly categorized into three types: directly providing emotion categories, predicting emotions from text, and extracting emotions from reference speech.
In the approach that directly provides emotion categories, C Cui et al. \cite{cui2021emovie} directly provide emotion labels, which are sent into the emotion predictor along with the results from the text encoder. The generated emotional information is then fed into the subsequent parts of the model.
In the approach that predicts emotions from text, Tu et al. \cite{tu2022contextual} argue that emotional labels depend on the contextual situation rather than specific tags. Therefore, they predict emotions by modeling the context.
In the approach that extracts emotions from reference audio, S. Oh et al. \cite{oh2019determination} proposed the Global Style Token (GST) method. This method encodes reference audio and combines it with global emotion labels through attention mechanisms, using the result as an emotional embedding. R. Liu et al. \cite{liu2021reinforcement} employed a similar approach, adding emotion recognition for adversarial training to improve the accuracy of label generation. Additionally, D. Min et al. \cite{min2021meta} directly encoded the style of the reference audio and fed the result into the model. Kang et al. \cite{kang2023grad} utilized a diffusion model for speech synthesis, incorporating the encoded result into the diffusion steps.

Despite the outstanding emotional expression achieved by these methods, their inclusion of emotion information during training limits the inference stage to a fixed number of emotion categories, resulting in insufficient scalability and flexibility.
In the current context of the booming development of AIGC technology, fine-tuning large models with all parameters requires a significant amount of computational cost. Therefore, the LoRA \cite{hu2021LoRA} plugin method has emerged. It records new knowledge by freezing the original parameters of the model and multiplying two low-rank matrices. When facing different downstream segmentation tasks, LoRA can train multiple sets of plugins to achieve plug-and-play simultaneously.

Due to the low computational cost and flexibility of LoRA, many related works have emerged.
V Shah et al. \cite{shah2023zipLoRA} used LoRA to independently learn the style and content of images, achieving arbitrary combinations of style and content.
CP Hsieh et al. \cite{hsieh2022adapter} inserted LoRA into the attention mechanism of the FastPitch \cite{lancucki2021fastpitch} model to learn the timbre differences between unseen and seen speakers, enabling the model to synthesize speech for new speakers.
Z Song et al. \cite{song2024LoRA} utilized LoRA to capture language-specific features, enabling multilingual speech recognition.
Additionally, there are some related variants.
Y Wang et al. \cite{wang2024residualtransformer} additionally introduced diagonal weight matrices to enhance the modeling capability of low-rank matrices.
S. Hayou et al. \cite{hayou2024LoRA+} set different learning rates for two low-rank matrices to improve model training efficiency.
L. Zhang et al. \cite{zhang2023LoRA} froze one of the two low-rank matrices and trained only the other matrix independently. This halved the number of parameters while maintaining performance comparable to standard LoRA techniques.

Although there are many applications and improvements of LoRA, most of them are limited to adjusting the condition part of the model, which requires the model to pre-introduce control information. This actually limits the potential for further expansion of the model's functionality. Therefore, we propose the Exploring Efficient and Extensible LoRA Integration in Emotional Text-to-Speech (EELE) method.
We use the VITS2 model \cite{kong2023vits2} as our baseline, training the model with neutral speech data.
Subsequently, we added LoRA to different modules within the model and fine-tuned it with emotional speech data, using subjective and objective experiments to verify and test the impact of adding LoRA at different positions. We also trained a separate set of LoRA for each emotion to test LoRA's ability to learn different emotions. After endowing the neutral model with emotional generation capabilities, we further explore the impact of the rank size of LoRA and the difference compared to directly fine-tuning the entire model.
The main contributions in this paper are as follows:
\begin{itemize}
\item We have explored a more flexible and scalable approach for emotion adaptation in speech models.
\item Through experiments, we tested the impact of deploying the LoRA module at different positions on emotional synthesis and validated the effectiveness of the proposed method.
\item We tested and validated the impact of rank size on LoRA's emotional learning ability and compared the optimal scheme of adding LoRA with the approach of fine-tuning the entire model.
\end{itemize}

\section{Method}
This section mainly introduces our various attempts to add emotional generation capabilities to the neutral synthesis model using LoRA, as well as the models and methods we used.

We explored using the current mainstream VITS2 model \cite{kong2023vits2}. We added LoRA modules to the model using Microsoft's open-source LoRAlib package \cite{hu2022LoRA}. We experimented with eight different combinations of adding LoRA, as shown in Figure \ref{all_try}.

The VITS2 \cite{kong2023vits2} model utilizes GAN \cite{creswell2018generative} mechanisms during training. Since the discriminator does not participate in the inference process, its parameters are not frozen during fine-tuning. Similarly, the speech encoder also follows this approach.

\subsection{Where to add: Exploring the deployment of LoRA}
A text encoder is responsible for extracting text features. A stochastic duration predictor estimates duration features from these text features. The projection layer determines the sampling distribution. The flow layer converts the sampling results into acoustic features, and the decoder transforms these acoustic features into a speech waveform.

From the functions of these modules, it can be seen that the text encoder and projection layer are mainly responsible for front-end modeling, the flow layer and decoder are responsible for back-end modeling, and the duration predictor is responsible for modeling alignment information. So when adding LoRA to the model, it is mainly divided into these three categories for design. 

We integrated LoRA in eight different ways, as shown in Figure \ref{all_try}, to explore whether emotion is more suitable for modeling in the front-end text features or in the back-end acoustic features. Additionally, we investigated the impact of duration alignment information on emotional expression.

\subsection{How to add: Implantation of LORA in TTS model}
\begin{figure}[h]
  \centering
  \includegraphics[width=\linewidth]{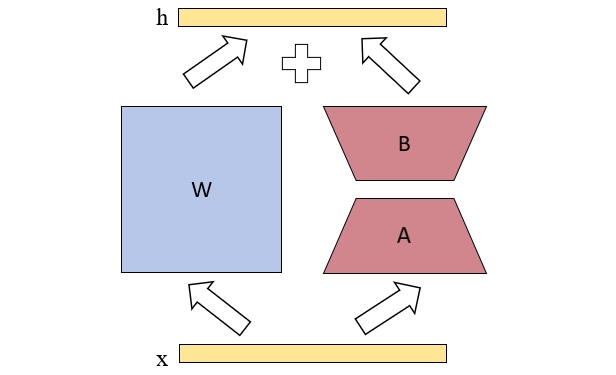}
  \caption{LoRA adds low-rank matrices to each layer.In the diagram, the blue \( W \) represents the frozen original model parameters, the red \( A \) and \( B \) represent the low-rank matrices, and the yellow \( x \) and \( h \) represent the layer's input and output, respectively.}
  \label{LoRA}
\end{figure}

In the VITS2 model \cite{kong2023vits2}, linear layers and 1D convolutional layers are primarily used. The LoRAlib package \cite{hu2022LoRA} conveniently includes methods for adding LoRA to these types of layers. After importing the package, we added low-rank matrices to each layer of the corresponding modules according to the eight different schemes, as illustrated in Figure \ref{LoRA}.

\begin{table*}[]
\centering
\caption{Subjective emotion recognition results}
\label{sub}
\begin{tabular}{l|c|c|c|c|c|c|c|c|c}
\hline
Emotion  & \textit{tts} & \textit{a} & \textit{b} & \textit{c} & \textit{d}    & \textit{e}    & \textit{f} & \textit{g}    & \textit{h} \\ \hline
Angry    & 0.64         & 0.63       & 0.76       & 0.77       & 0.81          & 0.65          & 0.65       & \textbf{0.88} & 0.78       \\ \hline
Happy    & 0.32         & 0.29       & 0.42       & 0.41       & 0.59          & 0.31          & 0.43       & \textbf{0.62} & 0.44       \\ \hline
Sad      & 0.17         & 0.18       & 0.28       & 0.20       & \textbf{0.46} & 0.31          & 0.27       & 0.45          & 0.33       \\ \hline
Supurise & 0.05         & 0.10       & 0.25       & 0.06       & 0.12          & \textbf{0.19} & 0.03       & 0.15          & 0.17       \\ \hline
\end{tabular}
\end{table*}

\begin{table*}[]
\centering
\caption{Objective emotion recognition results}
\label{Obj}
\begin{tabular}{l|c|c|c|c|c|c|c|c|c}
\hline
Emotion  & \textit{tts} & \textit{a} & \textit{b} & \textit{c} & \textit{d} & \textit{e} & \textit{f} & \textit{g}    & \textit{h} \\ \hline
Angry    & 0.7          & 0.67       & 0.78       & 0.71       & 0.82       & 0.68       & 0.67       & \textbf{0.86} & 0.83       \\ \hline
Happy    & 0.23         & 0.36       & 0.48       & 0.33       & 0.53       & 0.35       & 0.35       & \textbf{0.55} & 0.52       \\ \hline
Sad      & 0.12         & 0.22       & 0.33       & 0.23       & 0.39       & 0.21       & 0.23       & \textbf{0.42} & 0.31       \\ \hline
Supurise & 0.02         & 0.03       & 0.15       & 0.06       & 0.18       & 0.12       & 0.05       & \textbf{0.23} & 0.21       \\ \hline
\end{tabular}
\end{table*}

\section{Experiment Setup}
\subsection{DataSet}
We used the ESD \cite{zhou2022emotional} dataset and the VCTK \cite{Veaux2017CSTRVC} dataset.

The VCTK Corpus includes around 44 hours of speech data uttered by 110 English speakers with various accents. Each speaker reads out about 400 sentences, which were selected from a newspaper, the rainbow passage, and an elicitation paragraph used for the speech accent archive. We resampled all speeches to 16 kHz.

ESD is an Emotional Speech Database for voice conversion research. The ESD database consists of 350 parallel utterances spoken by 10 native English and 10 native Chinese speakers and covers 5 emotion categories (neutral, happy, angry, sad, and surprise). More than 29 hours of speech data were recorded in a controlled acoustic environment. The database is suitable for multi-speaker and cross-lingual emotional voice conversion studies. In this experiment, only the English data was used. We resampled all speeches to 16 kHz.

\subsection{Metric}
In the subjective experimental part, we hired 20 volunteers to conduct emotional evaluations of the experimental results and distinguish the emotional categories of speech. After the evaluation is completed, calculate the ratios recognized as corresponding emotions separately.

The objective experiment used a pre-trained model for emotion recognition \cite{enrique_hernández_calabrés_2024} to identify the experimental results. After the evaluation is completed, calculate the ratios recognized as corresponding emotions separately.

\subsection{Task}
In the subjective experiment, we hired 25 paid evaluators who had undergone professional training. They primarily assessed the eight methods of adding LoRA shown in Figure \ref{all_try} compared to the original without LoRA. For each emotion, they judged the proportion of sentences that were identified as expressing the corresponding emotion. Each evaluator assessed 300 sentences for each emotion.

In the objective evaluation, we used a pre-trained model instead of human judgment to perform the same type of experiment as in the subjective evaluation. Additionally, for the best-performing LoRA integration method, we tested the impact of matrix rank size and compared the emotional expression performance with that of fine-tuning.

\section{Experiment and Analysis}

\subsection{Subjective evaluation}
The results of the subjective experiment are shown in Table \ref{sub}.
Among them, $tts$ represents the evaluation results of the speech synthesized using a pre-trained neutral speech model without adding any LoRA plugins or conducting any fine-tuning.


Based on the results in the table, it can be observed that emotions with distinct characteristics, such as $angry$, often have high recognition rates. In contrast, more neutral emotions, such as $surprise$, are harder for people to perceive. The emotions $sad$ and $happy$ show relatively balanced performance. Among them, $g$ demonstrates notable expressiveness. We guess that emotional information is a speech feature, and therefore, assigning the learning of this knowledge to the model's acoustic modeling part can enhance its ability to model emotions. Additionally, the duration of pronunciation also affects emotional expressiveness, as people often use the speed of speech as one of the factors in determining the emotional category.

\subsection{Objective evaluation}
\subsubsection{Emotion recognition results}
The results of the objective experiment are shown in Table \ref{Obj}.
Among them, $tts$ represents the evaluation results of the speech synthesized using a pre-trained neutral speech model without adding any LoRA plugins or conducting any fine-tuning.

The objective experiment indicates that $g$ performs the best, which is consistent with the findings of the subjective experiment.
Among them, the recognition rate for $angry$ is the highest, with more noticeable improvements observed for $happy$ and $sad$, while $surprise$ still maintains the lowest recognition rate.

\subsubsection{The impact of changes in r-value}
The impact of the r-value on emotional performance is shown in Table \ref{rrr}.
$r$ represents the size of the matrix rank, where smaller values of $r$ indicate fewer parameters.
We selected the best-performing $g$ and conducted four experiments with $r$ values of 2, 4, 8, and 16.

From the experimental results, we found that varying the r-value did not appear to affect emotional performance.
We have two hypotheses: (1) We trained LoRA separately for each emotion, and emotions contain relatively limited information, which can be learned with a small number of parameters. (2) The emotional fine-tuning data we used may not be sufficient, allowing LoRA to fully learn the emotional information with fewer parameters.

\begin{table}[]
\centering
\caption{The impact of changes in r-value}
\label{rrr}
\begin{tabular}{l|c|c|c|c}
\hline
Emotion  & r=2  & r=4  & r=8  & r=16 \\ \hline
Angry    & 0.86 & 0.85 & 0.85 & 0.86 \\ \hline
Happy    & 0.55 & 0.54 & 0.55 & 0.55 \\ \hline
Sad      & 0.41 & 0.43 & 0.41 & 0.42 \\ \hline
Supurise & 0.21 & 0.23 & 0.22 & 0.23 \\ \hline
\end{tabular}
\end{table}

\subsubsection{Compared with fine-tuning}
The primary goal of this section's experiment is to observe whether our method has an advantage over conventional fine-tuning. The experimental results are shown in Table \ref{compare}.

We also used $g$ as the method for adding LoRA, with an r-value of 16. We observed that, apart from $angry$ which showed the same performance, the recognition rates for the other three emotions were slightly lower compared to direct fine-tuning.
We speculate that the decoupling of information in the current synthesis model is not yet complete. As a result, the text feature modeling part also affects emotional performance. If we continue to add LoRA to the relevant parts, the performance should improve.
Although the current experimental results are slightly behind, considering the number of parameters, scalability, and plug-and-play convenience, the method of using LoRA to endow the model with emotional expression capabilities is still highly advantageous.
\begin{table}[]
\centering
\caption{Comparison between fine-tuning and our method}
\label{compare}
\begin{tabular}{l|c|c}
\hline
Emotion  & \textit{g}  & \textit{fine-tune} \\ \hline
Angry    & 0.86 & 0.86      \\ \hline
Happy    & 0.55 & \textbf{0.67}      \\ \hline
Sad      & 0.42 & \textbf{0.55}      \\ \hline
Surprise & 0.23 & \textbf{0.33}     \\ \hline
\end{tabular}
\end{table}

\section{Conclusion}
In this paper, we introduced the EELE method, using LoRA to enable neutral TTS models to generate emotional speech without requiring emotion-specific training. This plug-and-play solution proved effective, creating a flexible, scalable emotional TTS system. Evaluations confirmed its ability to enhance emotional expression, with applications in movies, games, and voice acting. The EELE method advances adaptable emotional TTS systems, contributing to AI-generated content, with future work focusing on refining LoRA integration and broadening emotional expression.

\section{Acknowledgements}
This work is supported by the National Natural Science Foundation of China (NSFC) (No.62101553, No.62306316, No.U21B20210, No. 62201571).

\bibliographystyle{IEEEtran}

\bibliography{LoRA-emo}

\begin{thebibliography}{10}
\providecommand{\url}[1]{#1}
\csname url@samestyle\endcsname
\providecommand{\newblock}{\relax}
\providecommand{\bibinfo}[2]{#2}
\providecommand{\BIBentrySTDinterwordspacing}{\spaceskip=0pt\relax}
\providecommand{\BIBentryALTinterwordstretchfactor}{4}
\providecommand{\BIBentryALTinterwordspacing}{\spaceskip=\fontdimen2\font plus
\BIBentryALTinterwordstretchfactor\fontdimen3\font minus
  \fontdimen4\font\relax}
\providecommand{\BIBforeignlanguage}[2]{{%
\expandafter\ifx\csname l@#1\endcsname\relax
\typeout{** WARNING: IEEEtran.bst: No hyphenation pattern has been}%
\typeout{** loaded for the language `#1'. Using the pattern for}%
\typeout{** the default language instead.}%
\else
\language=\csname l@#1\endcsname
\fi
#2}}
\providecommand{\BIBdecl}{\relax}
\BIBdecl

\bibitem{kim2021conditional}
J.~Kim, J.~Kong, and J.~Son, ``Conditional variational autoencoder with
  adversarial learning for end-to-end text-to-speech,'' in \emph{International
  Conference on Machine Learning}.\hskip 1em plus 0.5em minus 0.4em\relax PMLR,
  2021, pp. 5530--5540.

\bibitem{popov2021grad}
V.~Popov, I.~Vovk, V.~Gogoryan, T.~Sadekova, and M.~Kudinov, ``Grad-tts: A
  diffusion probabilistic model for text-to-speech,'' in \emph{International
  Conference on Machine Learning}.\hskip 1em plus 0.5em minus 0.4em\relax PMLR,
  2021, pp. 8599--8608.

\bibitem{cui2021emovie}
C.~Cui, Y.~Ren, J.~Liu, F.~Chen, R.~Huang, M.~Lei, and Z.~Zhao, ``Emovie: A
  mandarin emotion speech dataset with a simple emotional text-to-speech
  model,'' \emph{arXiv preprint arXiv:2106.09317}, 2021.

\bibitem{tu2022contextual}
J.~Tu, Z.~Cui, X.~Zhou, S.~Zheng, K.~Hu, J.~Fan, and C.~Zhou, ``Contextual
  expressive text-to-speech,'' \emph{arXiv preprint arXiv:2211.14548}, 2022.

\bibitem{oh2019determination}
S.~Oh, S.-Y. Um, I.~Jang, C.~H. Ahn, and H.-G. Kang, ``Determination of
  representative emotional style of speech based on k-means algorithm,''
  \emph{The Journal of the Acoustical Society of Korea}, vol.~38, no.~5, pp.
  614--620, 2019.

\bibitem{liu2021reinforcement}
R.~Liu, B.~Sisman, and H.~Li, ``Reinforcement learning for emotional
  text-to-speech synthesis with improved emotion discriminability,''
  \emph{arXiv preprint arXiv:2104.01408}, 2021.

\bibitem{min2021meta}
D.~Min, D.~B. Lee, E.~Yang, and S.~J. Hwang, ``Meta-stylespeech: Multi-speaker
  adaptive text-to-speech generation,'' in \emph{International Conference on
  Machine Learning}.\hskip 1em plus 0.5em minus 0.4em\relax PMLR, 2021, pp.
  7748--7759.

\bibitem{kang2023grad}
M.~Kang, D.~Min, and S.~J. Hwang, ``Grad-stylespeech: Any-speaker adaptive
  text-to-speech synthesis with diffusion models,'' in \emph{ICASSP 2023-2023
  IEEE International Conference on Acoustics, Speech and Signal Processing
  (ICASSP)}.\hskip 1em plus 0.5em minus 0.4em\relax IEEE, 2023, pp. 1--5.

\bibitem{hu2021LoRA}
E.~J. Hu, Y.~Shen, P.~Wallis, Z.~Allen-Zhu, Y.~Li, S.~Wang, L.~Wang, and
  W.~Chen, ``Lora: Low-rank adaptation of large language models,'' \emph{arXiv
  preprint arXiv:2106.09685}, 2021.

\bibitem{shah2023zipLoRA}
V.~Shah, N.~Ruiz, F.~Cole, E.~Lu, S.~Lazebnik, Y.~Li, and V.~Jampani,
  ``Ziplora: Any subject in any style by effectively merging loras,''
  \emph{arXiv preprint arXiv:2311.13600}, 2023.

\bibitem{hsieh2022adapter}
C.-P. Hsieh, S.~Ghosh, and B.~Ginsburg, ``Adapter-based extension of
  multi-speaker text-to-speech model for new speakers,'' \emph{arXiv preprint
  arXiv:2211.00585}, 2022.

\bibitem{lancucki2021fastpitch}
A.~{\L}a{\'n}cucki, ``Fastpitch: Parallel text-to-speech with pitch
  prediction,'' in \emph{ICASSP 2021-2021 IEEE International Conference on
  Acoustics, Speech and Signal Processing (ICASSP)}.\hskip 1em plus 0.5em minus
  0.4em\relax IEEE, 2021, pp. 6588--6592.

\bibitem{song2024LoRA}
Z.~Song, J.~Zhuo, Y.~Yang, Z.~Ma, S.~Zhang, and X.~Chen, ``Lora-whisper:
  Parameter-efficient and extensible multilingual asr,'' \emph{arXiv preprint
  arXiv:2406.06619}, 2024.

\bibitem{wang2024residualtransformer}
Y.~Wang and J.~Li, ``Residualtransformer: Residual low-rank learning with
  weight-sharing for transformer layers,'' in \emph{ICASSP 2024-2024 IEEE
  International Conference on Acoustics, Speech and Signal Processing
  (ICASSP)}.\hskip 1em plus 0.5em minus 0.4em\relax IEEE, 2024, pp.
  11\,161--11\,165.

\bibitem{hayou2024LoRA+}
S.~Hayou, N.~Ghosh, and B.~Yu, ``Lora+: Efficient low rank adaptation of large
  models,'' \emph{arXiv preprint arXiv:2402.12354}, 2024.

\bibitem{zhang2023LoRA}
L.~Zhang, L.~Zhang, S.~Shi, X.~Chu, and B.~Li, ``Lora-fa: Memory-efficient
  low-rank adaptation for large language models fine-tuning,'' \emph{arXiv
  preprint arXiv:2308.03303}, 2023.

\bibitem{kong2023vits2}
J.~Kong, J.~Park, B.~Kim, J.~Kim, D.~Kong, and S.~Kim, ``Vits2: Improving
  quality and efficiency of single-stage text-to-speech with adversarial
  learning and architecture design,'' \emph{arXiv preprint arXiv:2307.16430},
  2023.

\bibitem{hu2022LoRA}
\BIBentryALTinterwordspacing
E.~J. Hu, Y.~Shen, P.~Wallis, Z.~Allen-Zhu, Y.~Li, S.~Wang, L.~Wang, and
  W.~Chen, ``Lo{RA}: Low-rank adaptation of large language models,'' in
  \emph{International Conference on Learning Representations}, 2022. [Online].
  Available: \url{https://openreview.net/forum?id=nZeVKeeFYf9}
\BIBentrySTDinterwordspacing

\bibitem{creswell2018generative}
A.~Creswell, T.~White, V.~Dumoulin, K.~Arulkumaran, B.~Sengupta, and A.~A.
  Bharath, ``Generative adversarial networks: An overview,'' \emph{IEEE signal
  processing magazine}, vol.~35, no.~1, pp. 53--65, 2018.

\bibitem{zhou2022emotional}
K.~Zhou, B.~Sisman, R.~Liu, and H.~Li, ``Emotional voice conversion: Theory,
  databases and esd,'' \emph{Speech Communication}, vol. 137, pp. 1--18, 2022.

\bibitem{Veaux2017CSTRVC}
C.~Veaux, J.~Yamagishi, and K.~MacDonald, ``Cstr vctk corpus: English
  multi-speaker corpus for cstr voice cloning toolkit,'' 2017.

\bibitem{enrique_hernández_calabrés_2024}
\BIBentryALTinterwordspacing
{Enrique Hernández Calabrés},
  ``wav2vec2-lg-xlsr-en-speech-emotion-recognition (revision 17cf17c),'' 2024.
  [Online]. Available:
  \url{https://huggingface.co/ehcalabres/wav2vec2-lg-xlsr-en-speech-emotion-recognition}
\BIBentrySTDinterwordspacing

\end{thebibliography}


\end{document}